\documentclass{mem}

\usepackage{natbib}
\usepackage{txfonts}
\usepackage{balance}
\usepackage{graphicx}
\usepackage[a4paper]{hyperref}
\idline{00}{183}

\def\kms{${\rm\,km\,s^{-1}}$}

\begin{document}

\title{Pattern speeds in the Milky Way}

\subtitle{}

\author{O. Gerhard}

\offprints{Ortwin Gerhard;\\ \email{gerhard@mpe.mpg.de}}

\institute{Max-Planck-Institut f\"{u}r extraterrestrische Physik,
  Garching, Germany}

\authorrunning{Gerhard}

\titlerunning{Pattern speeds in the Milky Way}

\abstract{

A brief review is given of different methods used to determine the
pattern speeds of the Galactic bar and spiral arms. The Galactic bar
rotates rapidly, with corotation about halfway between the Galactic
center and the Sun, and outer Lindblad resonance not far from the
solar orbit, $R_0$. The Galactic spiral arms currently rotate with a
distinctly slower pattern speed, such that corotation is just outside
$R_0$. Both structures therefore seem dynamically decoupled.

\keywords{Galaxy: structure -- Galaxy: kinematics and dynamics --
  Galaxy: fundamental parameters -- Galaxy: disk -- Galaxy: bulge --
  Galaxies: spiral} } 

\maketitle{}

\section{Introduction}

The Milky Way is a barred spiral galaxy with a boxy barred bulge
\citep[e.g.,][]{Dwek+95,BGS97}, extending to $\sim\,$2 kpc, and an
in-plane bar, reaching to $\sim\,$4 kpc
\citep[e.g.,][]{Benjamin+05,Cabrera-Lavers+07}.  The disk is probably
of Freeman type II, with an exponential profile outside $\sim\,$4 kpc
\citep{LC+04,Benjamin+05}, and a centrally flat or decreasing profile
inside this radius. The Galaxy probably has a four-armed spiral
pattern in the gas and young stars, but only two of these may be
present in the density distribution of old stars \citep[see
  e.g.,][]{Drimmel00,Martos+04}.

\section{Pattern speed of the Galactic bar}

The most direct determination of the bar's pattern speed has been
through applying a modified version of the Tremaine-Weinberg
continuity argument to a sample of $\sim\,$250 OH/IR stars in the
inner Galaxy \citep{DGS02}. The quantity actually measured is the
difference between the pattern rotation velocity and the circular
velocity at the local standard of rest (LSR). This depends on the
Galactic constants $R_0$, $V_0$ and the peculiar radial velocity
$u_{\rm LSR}$ of the LSR. The result is sensitive to $u_{\rm LSR}$;
however, HI absorption measurements show that the HI gas between the
Sun and the Galactic center moves at a common radial velocity
$-0.23\pm0.06$ \kms\ \citep{RS80}, so the most natural assumption is
that $u_{\rm LSR}$ is zero to this level.  \citet{DGS02} applied the
method to a subsample of $\sim\,250$ sufficiently bright and
long-lived OH/IR stars \citep[see][]{Sevenster02} to ensure an
approximately complete and relaxed sample.  The resulting value of the
pattern speed is $\Omega_p=59\pm5\pm10$ (sys) $\rm
km\,s^{-1}\,kpc^{-1}$ for $R_0=8$ kpc, $V_0=220$ \kms , and for other
values of the Galactic constants, $\Omega_p= (V_0/R_0) +
31.5\pm5\pm10$ (sys) $\rm km\,s^{-1}\,kpc^{-1}$ $-(18/R_0)\,u_{\rm
  LSR}$. The main part of the signal comes from stars at low $b$ and
around $l\sim\,30^\circ$, i.e., probably disk stars coupled to the
bar, perhaps near inner spiral arm tangent points or in an inner ring
around the bar \citep[see][]{SevKal01}.

More frequently, the pattern speed of the bar has been estimated from
comparing the gas flow in hydrodynamic simulations with the observed
Galactic CO and HI $lv$-diagrams. These simulations generally
reproduce a number of characteristic features in the $lv$-plot very
well, but none reproduces all the observed features. Thus the derived
pattern speeds vary somewhat.  \citet{EG99} obtain $\sim60$ $\rm
km\,s^{-1}\,kpc^{-1}$ (for `standard' $R_0=8$ kpc, $V_0=220$ \kms)
from placing the corotation radius $R_{\rm CR}$ outside the 3 kpc arm
and inside the molecular ring, and matching mainly to the spiral arm
tangents; \citet{Fux99} obtains $\sim50$ $\rm km\,s^{-1}\,kpc^{-1}$
($R_{\rm CR}=4-4.5$ kpc) from a comparison to several reference
features in the CO $lv$-plot; \citet{Weiner99} obtain $42$ $\rm
km\,s^{-1}\,kpc^{-1}$ ($R_{\rm CR}=5.0$ kpc) from matching the extreme
HI velocity contour; \citet{BEG03} obtain $55-65$ $\rm
km\,s^{-1}\,kpc^{-1}$ ($R_{\rm CR}=3.4\pm0.3$ kpc) from models with
separate bar and spiral pattern speeds and matching the spiral arm
ridges in the CO emission and the positions of molecular clouds and
HII region in the $lv$-diagram; \citet{RodrComb08} obtain $30-40$ $\rm
km\,s^{-1}\,kpc^{-1}$ and $R_{\rm CR}=5-7$ kpc from models with a
second nuclear bar and matching to the Galactic spiral arm pattern.
From these results, we may take as combined estimate for the bar
pattern speed from gas dynamics $\Omega_p=52\pm10$ $\rm
km\,s^{-1}\,kpc^{-1}$ or $R_{\rm CR}=3.5-5.0$ kpc (for `standard'
$R_0=8$ kpc, $V_0=220$ \kms; roughly $R_{\rm CR} \propto R_0$).

A third estimate for the bar pattern speed comes from determining the
length of the bar and assuming that, like in external galaxies the
Galactic bar is a fast bar, i.e., ${\cal R}=R_{\rm CR}/R_{\rm
  B}=1.2\pm0.2$ \citep{Aguerri+03}. The length of the NIR bar from
COBE is $R_{\rm B}\simeq 3.5$ kpc \citep{BGS97, BG02}, whereas the
length of the `long bar' from starcounts is $R_{\rm B}\simeq 4.0$ kpc
\citep{Benjamin+05,Cabrera-Lavers+07}. This results in a rather wide
range of $R_{\rm CR}=3.5-5.6$ kpc or $\Omega_{\rm b}\sim35-60$ $\rm
km\,s^{-1}\,kpc^{-1}$.

A final method is based on the interpretation of star streams observed
in the stellar velocity distribution function (VDF) in the solar
neighborhood as due to resonant orbit families near the outer
Lindblad resonance (OLR) of the Galactic bar.  Near the OLR there are
two elongated families of periodic orbits (anti-aligned inside and
aligned outside OLR), so an observer located near the points in the
disk where these cross may see two stellar streams at different
velocities \citep{Kalnajs91}. The associated non-periodic orbits from
both families can generate two streams in observations from a range of
radii and bar angles. Using a series of backward integration test
particle simulations to match the observed VDF, \citet{Dehnen00}
estimates $\Omega_{\rm b}=1.85\pm0.15 V_0/R_0$ ($51\pm4$ $\rm
km\,s^{-1}\,kpc^{-1}$ for the `standard' $R_0=8$ kpc and
$V_0=220$\kms). \citet{MuehlbDehn03} expanded on this model and show
that if the OLR of the bar lies slightly inside the solar circle and
the Sun lags the bar by $\sim 20^\circ$, three observed facts can be
explained: the lack of significant radial motion of the LSR, the
vertex deviation of $\sim 10^\circ$ for the old stars, and the
observed ratio of velocity ellipsoid
$(\sigma_2/\sigma_1)^2=0.42<0.5$. \citet{Minchev+07} show that the
observed value of the Oort $C$ constant as a function of velocity
dispersion can also be explained in this model, if $\Omega_{\rm
  b}=1.87\pm0.04 V_0/R_0$ ($51.5\pm1.5$ $\rm km\,s^{-1}\,kpc^{-1}$ for
`standard' $R_0$, $V_0$). This work suggests that the Galactic bar is
important for the velocity distribution near the
Sun. \citet{Chakrabarty07} agrees with this conclusion but concludes
that spiral arm perturbations need to be included, and criticizes the
backward integration simulations of \citet{Dehnen00}. Her best
estimate for bar corotation and pattern speed are $R_0/R_{\rm
  CR}\simeq 2.1\pm0.1$ and $\Omega_{\rm b}\simeq 57.5\pm 5$ $\rm
km\,s^{-1}\,kpc^{-1}$.

Rather than considering the response of disk particles to a bar-like
perturbation, \citet{Fux00} analyzes a fully self-consistent N-body
simulation of a barred galaxy scaled to the Milky Way. This simulation
shows multiple, time-dependent streams in many places in the disk, and
in particular often displays a Hercules-like stream outside
corotation. This stream is made of particles on `hot' orbits with
Jacobi energy just above its value at the 1-2-Lagrange points. While
this explanation of the Hercules stream is different from the OLR
scattering mechanism, it also places the OLR of the bar (at 7.7 kpc)
near $R_0$ (assumed 8 kpc in his model). Stellar kinematic data over
larger portions of the Galactic disk will be needed to identify the
correct mechanism and redetermine the final range of $\Omega_{\rm b}$.

\section{Spiral arm pattern speed}

It is not a priori clear whether the rotation of the Galactic spiral
pattern can be described by a constant, single pattern speed, and for
how long this approximation is valid. However, a useful first step is
to see whether this assumption is consistent with available data.

Here the most direct method relies on the birthplaces of the observed
open clusters. These are obtained from their current locations in the
disk by rotating them backwards in time along their orbits according
to their known ages, using a model for the local circular speed in the
disk. If open clusters are born in spiral arms, the distribution of
birthplaces for some age bin should be spiral-like, and by comparing
the spiral patterns obtained from different age bins, the rotation
rate of the pattern can be estimated.  \citet{DiasLepine05} did this
(i) by simple backward circular rotation for a sample of 599 clusters,
and (ii) by integrating the full orbits backwards for a sample of 212
clusters with radial velocities, proper motions, distances and
ages. They find that indeed most open clusters are born in spiral
arms, that the spiral arms approximately rotate like a rigid body, and
that $\Omega_{\rm sp}=24-26$ $\rm km\,s^{-1}\,kpc^{-1}$, so $R_{\rm
  CR,sp}=(1.06\pm 0.08) R_0$.

\citet{BEG03} computed gas flow models in realistic Galactic
potentials with different pattern speeds for the bar and spiral
pattern. Compared to single pattern speed models where the corotation
radius forms a rigid barrier for the gas flow, gas may flow inwards
through $R_{\rm CR}$ along arms passing through this region in models
with two pattern speeds. In the $lv$-plot, two pattern speeds models
therefore show regions with low gas content at radii around the bar's
corotation radius. Similar voids are present in the observed
$lv$-diagram. While this argues for a lower pattern speed for the
spiral arms than for the bar, the gas flow models have so far not been
accurate enough for this signature to reliably constrain the second
pattern speed. In the \citet{BEG03} simulations, both models with
$\Omega_{\rm sp}=20$ and $40$ $\rm km\,s^{-1}\,kpc^{-1}$ are
consistent with the data.

\citet{Martos+04} considered the self-consistent response of the
Galactic disk to the two-armed $K$-band spiral pattern proposed by
\citet{Drimmel00}, as a function of its pattern rotation $\Omega_{\rm
  sp}$. They find that dynamical consistency is sensitive to
$\Omega_{\rm sp}$, with the best results for $\Omega_{\rm sp}=20$ $\rm
km\,s^{-1}\,kpc^{-1}$. Using gas-dynamical simulations, they find that
the gaseous response to this two-armed pattern is a pattern of four
arms which resembles the Galactic pattern inferred from the tangent
points seen in various tracers \citep[e.g.][]{EG99}.

A relatively large literature exists on determining the spiral arm
pattern speed by fitting a kinematic model to the kinematics of OB
stars and Cepheids. The fitted models allow for solar motion, Galactic
rotation including values for $R_0$, $V_0$ and Oort constants, and the
kinematic response to an assumed spiral arm
perturbation. $\Omega_{\rm sp}$ and other free parameters are obtained
from the fit. Older determinations from OB and Cepheid stars
\citep[see][]{Fernandez+01} give $\Omega_{\rm sp}=20-30$ $\rm
km\,s^{-1}\,kpc^{-1}$. One study is by \citep{MishZen99} who model
Cepheid radial and Hipparcos proper motions, obtaining $\Omega_{\rm
  sp}-\Omega_0=0.4-2.2$ $\rm km\,s^{-1}\,kpc^{-1}$ and $R_{\rm
  CR,sp}-R_0=0.1-0.4$ kpc, for $\Omega_0=27.5$ $\rm
km\,s^{-1}\,kpc^{-1}$ and $R_0=8$ kpc. \citet{Fernandez+01} fit a
kinematic model to the Hipparcos O, B, and Cepheid velocities, finding
$\Omega_{\rm sp}=30$ $\rm km\,s^{-1}\,kpc^{-1}$.  \citet{Lepine+01}
investigate a superposition of $m=2$ and $m=4$ modes for the Galactic
spiral pattern and fit parameters of this model and the Galactic
constants to the Cepheid kinematics. They obtain similar values for
the pattern speeds of both modes, with $\Omega_{{\rm sp,}
  m=2}-\Omega_0=0.15\pm0.5$ $\rm km\,s^{-1}\,kpc^{-1}$ and
$\Omega_{{\rm sp,} m=4}-\Omega_0=0.18\pm0.1$ $\rm
km\,s^{-1}\,kpc^{-1}$, indicating that the Sun is within 0.2 kpc of
the corotation resonance of the pattern.

\citet{QuillMinch05} investigated the effect of a rotating spiral
pattern on the velocity distribution of old stars in the solar
neighborhood. They find that two families of orbits can be caused by
spiral density waves if the Sun is near the inner 4:1 resonance. This
gives $\Omega_{\rm sp}\sim 18$ $\rm km\,s^{-1}\,kpc^{-1}$ but the
match of the observed VDF near the Sun is not as good as in the
bar-driven models.  \citet{Chakrabarty07} simulated the effects of the
bar {\sl and} spiral arms on the VDF. She did not find a clear
best-fit model, but constrains $\Omega_{\rm b}\simeq 57.5\pm5$ $\rm
km\,s^{-1}\,kpc^{-1}$ and $\Omega_{\rm sp}\simeq (17-28)$ $\rm
km\,s^{-1}\,kpc^{-1}$.

\section{Conclusion}

The Galactic bar rotates rapidly, with corotation about halfway
between the Galactic center and the Sun, and OLR not far from the
solar orbit, $R_0$. To recapitulate, for $R_0=8$ kpc, $V_0=220$ \kms:
direct determination favors a fast pattern speed for the bar,
$\Omega_{\rm b}=59\pm5\pm10$ (sys) $\rm km\,s^{-1}\,kpc^{-1}$.
Hydrodynamic models from a number of papers give best fits to CO
$lv$-plots for $\Omega_{\rm b}\simeq 52\pm10$ $\rm
km\,s^{-1}\,kpc^{-1}$.  The velocity distribution of old stars in the
solar neighborhood seems influenced most by the bar, and somewhat by
the spiral arms. Modeling this gives $\Omega_{\rm b}\simeq50-60$ $\rm
km\,s^{-1}\,kpc^{-1}$. Taking all constraints together, the most
likely range is $\Omega_{\rm b}\simeq50-60$ $\rm
km\,s^{-1}\,kpc^{-1}$, corresponding to bar corotation at $R_{\rm
  CR}\simeq 3.5-4.5$ kpc.

The Galactic spiral arms rotate with a distinctly slower pattern
speed.  Open cluster birthplace analysis and the velocity field of
nearby young stars indicate that the current corotation of the spiral
pattern is just outside $R_0$. These tracers cover the last
$10^7-10^8$ yr, and result in $\Omega_{\rm sp}\simeq 25\pm2$ $\rm
km\,s^{-1}\,kpc^{-1}$.  Investigations of the stellar velocity
distribution in the solar neighborhood allow a wider range,
$\Omega_{\rm sp}\simeq (17-28)$ $\rm km\,s^{-1}\,kpc^{-1}$.  This
corresponds to a backward time-scale more like $\sim 10^9$ yr.
Hydrodynamic models also favor a second, slower pattern speed for the
spiral arms in the disk. The Galactic bar and spiral pattern thus seem
to be dynamically decoupled.

\bibliographystyle{aa}

\end{document}